\documentclass[a4paper,11pt,fleqn]{article}
\usepackage{latexsym}

\author{Sudhir R. Jain \footnote{srjain@apsara.barc.ernet.in}\\
Theoretical Physics Division, Bhabha Atomic Research Centre \\
Trombay, Mumbai 400 085, India}

\title{ Dissipation in finite Fermi systems }
\date{}
\begin{document}
\maketitle

\begin{abstract}
We present a systematic theory of dissipation in finite Fermi systems.
This theory is based on the application of periodic orbit theory to linear
response  of many-body systems. We concentrate only on the mesoscopic aspect
of the phenomena wherein a many-body system can be reduced to a single-body
in an effective mean-field. We obtain semiclassical periodic-orbit corrections
on top of the two-time correlation function for the rate of energy dissipation.
We show that this energy dissipation is irreversible on an observational time-
scale. To do so, we derive a generalised Smoluchowski equation for the energy
distribution in the quantal domain. Employing the Weyl-Wigner expansion, we also
obtain the equation governing the evolution of energy distribution in the
combination of semiclassical and adiabatic approximations. Further, we show how
the periodic orbit corrections are related to geometric phase acquired by single-
particle wavefunction as it evolves with the slow, time-varying mean-field.
We present our results for the important case of mixed dynamics also. We obtain
random-matrix results for response functions. We incorpoate chaos in the underlying
classical system by writing an ansatz for a generic wavefunction which leads
to various central results of equilibrium statistical mechanics.
This new formalism is extended here to include dissipation.
Finally, we present an expression for the viscosity tensor encountered
in nuclear fission in terms of periodic orbits of single particle in
an adiabatically deforming nucleus.

\end{abstract}

\frenchspacing


\frenchspacing


\newpage

\noindent
{\bf 1. Introduction}\\

Nuclei and metallic clusters support a number of collective excitations
which are well-known and well-studied \cite{bertsch-broglia,bromley}.
Damping of these collective excitations is not well understood. Dissipation
related to chaotic motion of a single particle in an effective mean-field has
been proposed and pursued for last two decades \cite{gross,swiatecki}.
We are interested in the mesoscopic aspect of this phenomenon where the
mean free path of a single particle is of the order of the size of the system.
For instance, the frequency at which an isovector giant dipole resonance is
excited in a nucleus is such that the corresponding wavelength is larger than
the nuclear size. In the standard parlance of nuclear physics, this is referred
to as one-body dissipation. In this paper, we present a systematic semiclassical
theory of one-body dissipation. Since it is well-known that several spectral
properties of chaotic quantum systems, nuclei etc. are well represented by
random matrix theory, we present results where random matrix theory is conjointly
taken with linear response theory.

This work is inspired by some recent developments where some of the most
important results of equilibrium statistical mechanics have been derived, starting
from chaos in the single-particle motion in gases. Due to presence  of chaos,
the eigenfunctions of the corresponding quantum system become irregular, where
irregularity is quantified by some correlations.
For instance, spatial correlation function
of an irregular eigenfunction
is given by a Bessel function \cite{berry,srednicki}.
The eigenfunctions can be modelled by a random superposition of plane waves.
Remarkably, from this ansatz, without thermal reservoir, all the momentum
distributions (viz., Maxwell-Boltzmann, Fermi-Dirac, Bose-Einstein) have been
re-derived by Srednicki \cite{srednicki}.
This approach  plays a very important role in two dimensions where
some of the standard methods of statistical mechanics fail.
For the two-dimensional case,
the momentum distribution was found \cite{jain-alonso} for an anyon gas
upto ${\mathcal O}(\hbar ^2)$ .
It was also shown \cite{alonso-jain} that the ansatz is consistent with
random matrix theory. Employing a random matrix argument, the  second law of
thermodynamics could  also be arrived at by Jain and Alonso \cite{
jain-alonso}.
In this work, we attempt to extend this framework to include one-body dissipation.

Let us recall that for a nucleus participating in a reaction, the
time-scale is $\sim 10^{-20}$s or even shorter. It has been qualitatively argued
\cite{hw} that in this situation, a nucleus is an isolated system. Thus,
we must arrive at the notion of dissipation for an isolated quantum system
since we have already succeeded in arguing for thermalisation of isolated
quantum many-body systems \cite{srednicki}.

A  formal theory of one-body dissipation was developed in  classic works
where linear response techniques were used \cite{koonin-randrup,hofmann}. Our
considerations also rest on linear response theory. Beginning from linear response
theory, we present a semiclassical expression for response function. For
sake of comparison, we also present random matrix  expression. From this,
we obtain the rate of energy dissipation. Clearly, if there is energy
dissipation, the energy distribution must evolve irreversibly in the quantal
problem. This becomes an important issue as there can be no irreversibility
or dissipation in an isolated quantal system that supports only pure point spectrum
\cite{berry-robbins}. Employing multiple time-scale analysis, we show that
for sufficiently long times, there is irreversibility if the spectrum is
coarse-grained (e.g., by a random matrix ensemble average).
Beginning from the von Neumann
equation for the density operator, we show that the energy distribution
obeys a generalised Smoluchowski equation. When semiclassical and adiabatic
approximations are put together, it is a highly singular situation. We
perform a double expansion, one in the small adiabaticity parameter,
$\epsilon $, and the other in $\hbar $. We then obtain the equation for
energy distribution valid upto times of order $\epsilon ^{-1}$, and
upto order $\hbar ^2$. With this derivation, we have thus found the
solution of the following general problem. If a quantal system is evolving
adiabatically, the energy levels of the system will evolve in a complicated
manner \cite{footnote1}. As the levels come much closer than the mean level
spacing at a particular epoch of evolution, the non-adiabatic transitions
become more probable. This can be understood by employing degenerate
perturbation theory. Thus, if we distribute some particles
initially (at time, $t = 0$) in accordance with certain law (e.g.,
the Boltzmann law), the final distribution will be very different. It
has been thought that the evolution of the energy distribution must be
diffusive. Although the guess is qualitatively correct, the equation
is not the Smoluchowski equation but another equation which we have
called as generalised Smoluchowski equation \cite{footnote2}.
This general problem is encountered in many fields of physics and
chemistry.

Subsequently, we present relations connecting single-particle properties
and bulk properties. We show that the geometric phase acquired by a
single-particle wavefunction due to adiabatically vibrating mean-field
is related to the response function. Although this result has been obtained
earlier \cite{jain-pati}, we briefly discuss here to put it in proper
perspective.

All the above results assume that single-particle dynamics is
chaotic. In realistic systems, all the symmetries are not broken.
Thus the dynamics is usually mixed, i.e., there are
stable islands embedded in stochastic sea. We obtain the results for this
case which are based on trace formula for the case of partially broken
symmetry obtained by Creagh \cite{creagh}.

The response function should depend on the
shape of the cavity or the mean-field potential. In a relatively recent
attempt, shape dependence was incorporated by multiplying by a factor,
$\overline{\mu }$, the average fraction of trajectories which are chaotic
when a uniform sampling is done \cite{pal}. This is based on some heuristic
arguments. It was shown that this chaos weighted wall formula is in a better
agreement with experimental data \cite{pal1}. This observation and knowing
that the spectral properties of non-integrable systems can be understood
using semiclassical arguments based on the trace formulae \cite{jain-parab},
we believe that the theory developed here is logical and natural.
Since it is known that the spectral properties of chaotic quantum
systems can be understood in terms of semiclassical trace formulae,
our approach is more general and system-specific.
We believe that the development presented here provides a better understanding
of the interplay between single-particle motion and collective motions in
a many-body quantum system.

\vskip 0.5 truecm

\noindent
{\bf 2. One-body dissipation and wall formula}\\

To understand damping in low-energy heavy-ion collisions, the
concept of one-body dissipation was introduced \cite{gross} where
dissipation results from the interaction of nucleons with the
time-varying mean-field. This involves a reduction of many-body system
to a single-particle in an effective mean-field. Although two-body
collisions are also important, due to Pauli exclusion principle, the
available phase space for such collisions is much lesser and the
one-body mechanism is expected to dominate in the mesoscopic regime.

Assuming that the total one-body Hamiltonian is $H({\bf r},{\bf p};t)$
where $({\bf r},{\bf p})$ are phase space coordinates of a single
particle. Splitting $H$ into a time-independent and a
time-dependent part, respectively $H_0({\bf r},{\bf p})$ and
$H_1({\bf r},{\bf p};t)$, under the assumptions of linear response
theory, the rate of energy dissipation is given by \cite{koonin-randrup},
\begin{eqnarray}
\dot{E} &=& - \int \frac{d{\bf r}d{\bf p}}{(2\pi \hbar)^3}\left[
\int_{0}^{\infty} dt' \dot{H_1}({\bf R}_0({\bf r},{\bf p};t'),
{\bf P}_0({\bf r},{\bf p};t');t)\right]\nonumber \\
& &\dot{H_1}({\bf r},{\bf p};t)
\frac{\partial f_0}{\partial H_0}({\bf r},{\bf p}),
\end{eqnarray}
where $f_0$ is the single-particle phase space distribution and
an overdot represents a time-derivative. The important quantity in (1)
is the two-time correlation function in the integrand.

The rate of energy dissipation for the case when the dynamics of a
single-particle is chaotic is given by the wall formula \cite{swiatecki} :
\begin{equation}
\dot{E}_{\mathrm{wall}} = \rho \overline{v} \int u^2(a)d^2a,
\end{equation}
where  $u(a)$ is the normal component of the surface velocity at the
point $a$ on the surface, $\rho $ and $\overline{v}$ are nuclear mass density
and average nucleon speed inside the nucleus.

The above expressions hold when the dynamics of a single particle is
fully chaotic. However, since some of the symmetries are not broken,
the dynamics is expected to be mixed rather than chaotic. For this case,
a scaled wall formula \cite{pal},
\begin{equation}
\dot{E} = \overline{\mu }\dot{E}_{\mathrm{wall}},
\end{equation}
has been proposed. The factor, $\overline{\mu}$, is the average fraction
of the trajectories which are chaotic when a uniform sampling of the
surface is done. The factor is fixed from the parameter appearing in the
Berry-Robnik distribution \cite{berry-robnik}, which gives the nearest-
neighbour level spacing distribution for systems with mixed phase space.
This way of incorporating the role of dynamics is empirical. In
our framework, we take care of the  dynamical features explicitly,
hence any role of dynamics enters explicitly. Importantly, since
we employ the trace formula, there is harmony between the dynamics and
the spectral statistics.

Let us now comment on the relation between chaos and dissipation. From
a number of studies, classical \cite{blocki} and quantal \cite{skalski},
it is  clear that if the dynamics is chaotic, there is dissipation.
If the
particle and wall motions are considered self-consistently, then it
has been numerically seen that the dynamics is chaotic \cite{baldo}.
In this case, the energy is gained irreversibly, leading to dissipation.
If the dynamics is regular
or integrable, the energy gained by a particle from the wall will
eventually be fed back to the wall, resulting in no dissipation.

The one-body mechanism is dominant for low-energy nucleus-nucleus
collisions or fission, i.e., in general,  to slow collective processes.
It is known that to simultaneously reproduce both the fission probability
and the pre-scission neutron multiplicity, one needs a shape-dependent
friction \cite{froebrich1,froebrich2}.
These works came up as an attempt to combine dynamics and thermodynamics
in a consistent way. The shape of the nucleus is reflected in the level density
and this is used, in turn, to construct entropy. At this point, whereas
these works in nuclear physics accept a stochastic description in working
with Langevin equation, we would persevere with the dynamical considerations.
We believe that friction or viscosity, if any, in these quantum systems will
have quantal signatures. For instance, the response function of a many-body
Fermi system is related to geometric phase acquired by a single-particle
eigenfunction as the system deforms \cite{jain-pati}.
These considerations need a
systematic theory where shape plays an important role. To do so, we
believe that it is important to employ the Gutzwiller \cite{gutzwiller}
trace formula and evolve a framework.
\vskip 0.5 truecm

\noindent
{\bf 3. Quantum chaos and equilibrium statistical mechanics}\\

The connection between chaos and equilibration or thermalization of
many-body systems has been always believed in.
The tenets of equilibrium statistical mechanics rest on the assumption
of ``molecular chaos" - an idea that goes back to Boltzmann \cite{klein}.
With the advancement of our understanding of the theory of dynamical
systems and its relation with statistical mechanics \cite{gaspard},
it has become important to extend this understanding to quantal many-body
systems. Only recently,  such a connection  has been established.
Since this connection and the present approach
is an underlying theme of this paper, we briefly  present the
status of our understanding for want of a better perspective.

Let us mention the numerical results on eigenfunctions of
systems whose classical counterparts  are chaotic.
\vskip 0.25 truecm
\noindent
({\bf N1}) The amplitude distribution of eigenfunctions are found to
agree with a Gaussian distribution - we call these states as the generic
ones \cite{mcdonald}.
Of course, we are not referring to eigenfunctions corresponding to
the ground state which is special. Nevertheless, this observation is more
general as the Gaussian distribution remains good even for pseudointegrable
(non-integrable and non-chaotic) systems like a $\pi /3$-rhombus billiard
\cite{jain1,jain2} where the Kolmogorov-Sinai entropy is zero.
\vskip 0.25 truecm
\noindent
({\bf N2}) The spatial correlation function,
\begin{equation}
C_{\alpha }({\vec s}) = {1 \over V} \int d{\bf r}
\psi _{\alpha }^{\star}\left( {\bf r}-\frac{{\bf s}}{2}\right)
\psi _{\alpha }\left( {\bf r}+\frac{{\bf s}}{2}\right),
\end{equation}
corresponding to an eigenfunction $\psi _{\alpha }({\bf r})$ is found to
agree with a (cylindrical) Bessel function \cite{berry},
where $V$ is volume. For two-dimensional billiards,
it has been found to agree with $J_0(k{\bf s}), k = \sqrt{2mE/\hbar ^2}$
\cite{mcdonald}.
\vskip 0.25 truecm
\noindent
({\bf N3}) The nodal lines are quite complicated, for pseudointegrable
as well as chaotic billiards \cite{mcdonald,jain2}.
\vskip 0.25 truecm

As far as eigenvalues or the energy levels are concerned, the measures
quantifying spectral statistics agree with the results known from random
matrix theory \cite{bohigas}. Even for pseudointegrable billiards, which are
non-integrable, the nearest-neighbour level spacing distribution can be
explained in terms of new ensembles \cite{gremaud}.

The numerical results ({\bf N1,N2}) can be explained if we assume that the
complicated eigenfunctions are written as a random superposition of plane
waves. This has been advocated by Berry \cite{berry} and written explicitly
by Srednicki \cite{srednicki}. For an $N$-particle system, the canonical pair of
coordinates are $({\bf X},{\bf P})$ where
${\bf X}=({\bf x}_1,{\bf x}_2,...,{\bf x}_N)$ and ${\bf P}=({\bf p}_1,{\bf p}_2,...,{\bf p}_N)$.
The energy eigenfunction
$\psi _{\alpha }({\bf X})$ is thus written as
\begin{equation}
\psi _{\alpha }({\bf X}) = N_{\alpha } \int d^{dN}{\bf P}
A_{\alpha}({\bf P})\delta (P^2-2mE_{\alpha})e^{\frac{i}{\hbar }{\bf X}.{\bf P}}
\end{equation}
for a $d$-dimensional system, where $N_{\alpha}$ is the normalization
constant and the amplitudes satisfy the two-point correlation function,
\begin{equation}
\left< A_{\alpha}^{\star}({\bf P})A_{\gamma}({\bf P}') \right>_{ME}
= \delta _{\alpha \gamma} \frac{\delta ^{dN}({\bf P}-{\bf P}')}{\delta
(P^2-P'^2)}.
\end{equation}
This correlator has been shown to be consistent with random matrix theory,
hence the appearance of a statistical average on the left side of the
equation ($ME$ stands for a matrix-ensemble average) \cite{alonso-jain}.
Beginning from this ansatz for the eigenfunction, the momentum distributions
for ideal gases have been shown to concur with  the well-known results (like
Maxwell-Boltzmann, Fermi-Dirac, Bose-Einstein, or fractional statistics)
by Srednicki \cite{srednicki}, and, Jain and Alonso \cite{jain-alonso}.
Even the first few virial coefficients have been calculated.
The Wigner distribution corresponding to an eigenfunction is microcanonical.
Thus, generically, corresponding to an eigenfunction, there is entire energy
surface in phase space.

An important question - in this theory, what is temperature ? Temperature
quantifies heat, and heat is defined as a mode of motion.
Even if there is no reservoir, one can still consistently use the kinetic
theory, and write for a system with $f$ freedoms at energy $E_{\alpha }$, a
temperature, $T_{\alpha }$, with $E_{\alpha }= \frac{1}{2}fkT_{\alpha }$,
where $k$ is the Boltzmann constant. As the average over the entire system
involves an averaging over the level density, the average temperature, $T$
will automatically get defined. This average temperature is the one we measure
as we obtain the correct momentum distributions with this
temperature playing  exactly the same role as the ``usual" temperature.

For the case of a many-body system like a nucleus where a nucleon trajectory
is chaotic, a typical eigenfunction  will be written as above. Thus, as the
system evolves and shape of the nucleus changes, the eigenfunction will
acquire a geometric phase (see Sec. 8). It is this phase which will eventually find a
relation with the absorptive part of generalised susceptibility \cite{jain-pati}.
\vskip 0.5 truecm

\noindent
{\bf 4. Semiclassical linear response}\\

To study the effect of external perturbation on a many-body system,
one can make usage of linear response theory \cite{kubo,kadanoff-martin,balescu}.
Also, one can imagine a fictitious external field to study collective
excitations as if they were generated due to a hypothetic field
\cite{hofmann}. In this section, we obtain an expression for generalised
susceptibility and the frequency-dependent response function (also called
as the polarisation propagator \cite{fetter}) in terms of periodic orbits
employing the Gutzwiller trace formula \cite{gutzwiller}. We assume that the
dynamics of a single particle is fully chaotic. As mentioned earlier, the
important case of mixed dynamics will be taken up in Section 6. The
assumption of chaos is consistent with some recent numerical inverstigations
for the case of nuclei \cite{baldo}.

The system is described by a Hamiltonian $\hat{H}$ which is disturbed by
a field, $F^{\mathrm{ext}}(t)$. The total Hamiltonian is
\begin{equation}
\hat{H}_T = \hat{H} - \hat{Q}F^{\mathrm{ext}}(t)
\end{equation}
where $\hat{Q}$ is an observable, an example could be magnetization in the
context of spin systems, or, an electric dipole operator in
photoabsorption experiments. The response function can be written as the
imaginary part of the dynamical susceptibility,
\begin{eqnarray}
\chi ^{\prime \prime}(t,t') &=& \frac{1}{2\hbar}\left< [\hat{Q}(t),
\hat{Q}(t')]\right> \nonumber \\
&=& \int \frac{d\omega}{2\pi} e^{-i\omega (t-t')}\tilde{\chi ^{\prime \prime}}(\omega ).
\end{eqnarray}
The angular brackets denote the expectation value and the square brackets
denote the commutator. Setting an initial time to zero and final time to $t$,
we can re-write
\begin{equation}
\chi ^{\prime \prime}(t) = \frac{1}{2\hbar} \left< [\hat{Q}(0),\hat{Q}(t)] \right>
\end{equation}
where
\begin{equation}
\hat{Q}(t) = e^{\frac{i}{\hbar}\hat{H}t}\hat{Q}e^{-\frac{i}{\hbar}\hat{H}t},
\end{equation}
$\left<...\right>$ denotes an average over the initial state
of the system which is, for instance, the thermal state wherein
\begin{equation}
\left<...\right> = \frac{1}{Z(\beta )}\mbox{tr~}e^{-\beta \hat{H}}(...)
\end{equation}
where  the temperature, $T = \frac{1}{k\beta}$.

Denoting the many-body eigenstates by $\Phi _n$ and one-body
eigenstates by $\phi _n$, and after reducing the description of the
many-body system to one-body, the propagator is
\begin{equation}
\tilde{\chi ^{\prime \prime}}(\omega ) = \sum_{a,b}
|\langle \phi _a |\hat{q}|\phi _b \rangle |^2 \Im
\frac{p^{FD}(\epsilon _a)-p^{FD}(\epsilon _b)
}{\hbar \omega - \epsilon _a + \epsilon _b + i0^{+}},
\end{equation}
where
\begin{equation}
p^{FD}(\epsilon _a) = \frac{1}{e^{\beta (\epsilon _a - \mu )}+1}
\end{equation}
denotes the Fermi-Dirac probability of occupation number with $\mu $
as the chemical potential.
Eq. (12) is the Fourier transform of the two-time correlation function.
In  (12), the operator $\hat{q}$ is the one-body operator which when
taken for each body and direct-summed gives us the operator, $\hat{Q}$.

Now that the susceptibility has been reduced to a 1-body expression,
we can treat the 1-body system semiclassically in terms of periodic
orbits in an effective Hartree-Fock (or some other mean-field)
Hamiltonian, $\hat{h}$. For subsequent convenience, let us re-write
(12) as
\begin{eqnarray}
\tilde{\chi ^{\prime \prime}}(\omega ) &=& \sum_{a,b}
|\langle \phi _a |\hat{q}|\phi _b \rangle |^2
[p^{FD}(\epsilon _a) - p^{FD}(\epsilon _a + \hbar \omega )]
\delta (\hbar \omega + \epsilon _a - \epsilon _b) \nonumber \\
&=& -\pi \sum_{n=1}^{\infty} \frac{(\hbar \omega )^n}{n!}
\sum_{a,b}\frac{\partial ^np^{FD}}{\partial \epsilon ^n}(\epsilon _a)
|\langle \phi _a |\hat{q}|\phi _b \rangle |^2
\delta (\hbar \omega + \epsilon _a - \epsilon _b) \nonumber \\
&\equiv & -\frac{1}{2\hbar} \sum_{n=1}^{\infty}\frac{(\hbar \omega )^n}{n!}
\tilde{f}_n(\omega ).
\end{eqnarray}
Here we have introduced the time correlation function $f_n(t)$ and its
Fourier transform, $\tilde{f}_n(\omega )$:
\begin{eqnarray}
f_n(t) &=& \mbox{tr~}\frac{\partial ^np^{FD}}{\partial \epsilon ^n}(\hat{h})
e^{{i \over \hbar}t\hat{h}}\hat{q}e^{-{i \over \hbar}t\hat{h}}\hat{q}; \nonumber \\
\tilde{f}_n(\omega ) &=& \int dt e^{i\omega t}f_n(t) \nonumber \\
&=& 2\pi \hbar   \sum_{a,b}\frac{\partial ^np^{FD}}{\partial \epsilon ^n}(\epsilon _a)
|\langle \phi _a |\hat{q}|\phi _b \rangle |^2
\delta (\hbar \omega + \epsilon _a - \epsilon _b).
\end{eqnarray}
In fact,
\begin{equation}
f_n(t) = \int d\epsilon
\frac{\partial ^np^{FD}}{\partial \epsilon ^n}(\epsilon )C(\epsilon , t)
\end{equation}
with
\begin{equation}
C(\epsilon , t) = \mbox{tr~}\delta (\epsilon - \hat{h})\hat{q}(t)\hat{q}(0).
\end{equation}
Finally,
the propagator is
\begin{equation}
\tilde{\chi ^{\prime \prime}}(\omega ) = -\pi \sum_{n=1}^{\infty}
\frac{(\hbar \omega )^n}{n!} \int \frac{dtd\epsilon }{2\pi \hbar }
e^{i\omega t}
\frac{\partial ^np^{FD}}{\partial \epsilon ^n}(\epsilon )C(\epsilon , t).
\end{equation}

The classical dynamics in the three-dimensional one-body effective potential
of a fermionic system will, in general, be chaotic. Now we apply the Gutzwiller
trace formula to obtain an expression for the propagator. For this, we need
the semiclassical expression for the time correlation, $C(\epsilon , t)$,
which we consider in the form :
\begin{eqnarray}
C(\epsilon , t) &=& \mbox{tr~}\delta (\epsilon - \hat{h})\hat{X} \nonumber \\
&=& \mbox{tr~}\delta (\epsilon - \hat{h})\hat{A} +
i   \mbox{tr~}\delta (\epsilon - \hat{h})\hat{B},
\end{eqnarray}
where $\hat{X}=\hat{X}(t)=\hat{q}(t)\hat{q}(0)$ is  one-body operator.
This operator is non-hermitian but can be decomposed as
$\hat{X}=\hat{A}+i\hat{B}$ in terms of two  hermitian operators :
\begin{equation}
\hat{A} = \frac{\hat{X}+\hat{X}^{\dagger}}{2},~~
\hat{B} = \frac{\hat{X}-\hat{X}^{\dagger}}{2i}.
\end{equation}
We need to re-express matrix elements of $\hat{A}$ (and $\hat{B}$)
overthe eigenstates of $\hat{h}$. These matrix elements can be obtained
by employing first-order perturbation theory for the perturbed
Hamiltonian $\hat{h}(\lambda )=\hat{h}+\lambda \hat{A}$. Assuming the
eigenvalue problem for $\hat{h}(\lambda )$ to be solved, the matrix elements
of $\hat{A}$ may thus be obtained in terms of derivatives of eigenvalues
of the perturbed Hamiltonian, $\epsilon _n(\lambda )$, with respect to
$\lambda $.

Each term of the correlation function $C(\epsilon , t)$ can be expressed as
\begin{eqnarray}
\mbox{tr~}\delta (\epsilon - \hat{h})\hat{A} &=& -{1 \over \pi}\Im
\mbox{tr~}\frac{\hat{A}}{\epsilon - \hat{h} + i0^+} \nonumber \\
&=& {1 \over \pi}\Im \mbox{tr~}\frac{\partial}{\partial \lambda}
\log (\epsilon - \hat{h} -\lambda \hat{A} + i0^+)\vline _{_{_{\lambda = 0}}}
\end{eqnarray}
On comparing with the identity,
\begin{eqnarray}
-{1 \over \pi}\frac{\partial}{\partial \epsilon} \Im \mbox{tr~}
\log (\epsilon - \hat{h} -\lambda \hat{A} + i0^+) &=& \mbox{tr~}
\delta (\epsilon - \hat{h} - \lambda \hat{A}) \nonumber \\
&=& \frac{\partial N}{\partial \epsilon}(\epsilon ;\lambda ),
\end{eqnarray}
we have
\begin{equation}
\mbox{tr~}\delta (\epsilon - \hat{h})\hat{A}
= - \frac{\partial N}{\partial \lambda}(\epsilon ;\lambda )
\vline _{_{_{\lambda = 0}}},
\end{equation}
where $N(\epsilon ;\lambda )$ is the cumulative density of levels for
the parametrised Hamiltonian $\hat{h}(\lambda )$.

For the case under consideration where the single-particle dynamics is
chaotic in an effective mean-field whereupon we can assume that the
periodic orbits are isolated and unstable, the cumulative density of levels
is given by the well-known expression due to Gutzwiller \cite{gutzwiller},
\begin{eqnarray}
N(\epsilon ;\lambda ) &=& \int \frac{d^fxd^fp}{(2\pi \hbar )^f}
\Theta [\epsilon - h_W(\lambda )] + {\mathcal O}(\hbar ^{-f+1}) \nonumber \\
&+& \sum_{p}\sum_{r=1}^{\infty } {1 \over r\pi}
\frac{\sin \left[ {r \over \hbar}S_p(\epsilon ;\lambda ) -
r{\pi \over 2}\nu _p\right]}{|\det ({\bf \sf m}_p^r(\lambda ) -
{\bf  \sf I})|^{1 \over 2}} + {\mathcal O}(\hbar ).
\end{eqnarray}
Here, ${\bf \sf m}_p^r(\lambda )$ is the monodromy matrix governing
the stability of the classical periodic orbits, $p$; $\nu _p$ denotes the
Maslov index of the trajectory, and $h_W$ is the Weyl symbol of the
Hamiltonian operator, $\hat{h}(\lambda )$.

Owing to (24), we need the derivative of the action w.r.t. the parameter,
$\lambda $ \cite{du}:
\begin{equation}
\frac{\partial S_p}{\partial \lambda} = -\oint dt
\frac{\partial h_W(\lambda )}{\partial \lambda} = -\oint _pdt A_W
\end{equation}
where $A_W$ is the Weyl symbol of $\hat{A}$. Thus,
\begin{eqnarray}
\mbox{tr~}\delta (\epsilon - \hat{h})\hat{X} &=& \int \frac{d^fxd^fp}{(2\pi \hbar )^f}
X_W\delta [\epsilon - h_{\mathrm{cl}} )] + {\mathcal O}(\hbar ^{-f+1}) \nonumber \\
&+& \frac{1}{\pi \hbar}\sum_{p}\sum_{r=1}^{\infty }
\frac{\cos \left[ {r \over \hbar}S_p -
r{\pi \over 2}\nu _p\right]}{|\det ({\bf \sf m}_p^r(\lambda ) -
{\bf  \sf I})|^{1 \over 2}}\oint_pdtX_W + {\mathcal O}(\hbar ^0),
\end{eqnarray}
where $X_W$ is the Weyl symbol of $\hat{X}$.

For our case where
\begin{equation}
\hat{X} = \exp \left({i \over \hbar}\hat{h}t\right)\hat{q}
\exp \left(-{i \over \hbar}\hat{h}t\right)\hat{q},
\end{equation}
the Weyl symbol can be written as \cite{balasz}
\begin{eqnarray}
X_W({\bf x},{\bf p}) &=& (e^{{i \over \hbar}\hat{h}t})_W
                        e^{{i \over 2    }\hbar\Lambda }
\bigg\{ q_W e^{{i \over 2    }\hbar\Lambda }
\bigg[
(e^{-{i \over \hbar}\hat{h}t})_W e^{{i \over 2    }\hbar\Lambda }q_W
\bigg] \bigg\} \nonumber \\
&=& \left[\exp (-\hat{{\mathcal L}}_{ cl}t)  \right]q_W
+ \mbox{Weyl-Wigner corrections} \nonumber \\
&=& q_W(t)q_W(0).
\end{eqnarray}

The operator, $\hat{{\mathcal L}}_{ cl}$ is the classical Liouvillian
operator defined by $\{h,. \}$, and the operator  $\hat{\Lambda }$ is defined by

\begin{equation}
{\hat{\Lambda }}= \frac{{\leftarrow \atop \partial }}{ \partial {\bf p}}.
 \frac{{\rightarrow \atop \partial}}{\partial {\bf x}}
          - \frac{{\leftarrow \atop \partial}}{\partial {\bf x}}.
            \frac{{\rightarrow \atop \partial}}{\partial {\bf p}},
\end{equation}

and $q_W(t)=q_W[\Phi ^t({\bf x},{\bf p})]$ with the Hamiltonian flow denoted
by $\Phi ^t$.

Finally, the correlation function is
\begin{eqnarray}
C(\epsilon , t) &=& \mbox{tr~}\delta (\epsilon - \hat{h})\hat{q}(t)\hat{q}(0)
\nonumber \\
&=& \int \frac{d^fxd^fp}{(2\pi \hbar )^f}
q_W\left[\exp (-\hat{{\mathcal L}}_{ cl}t)  \right]q_W
\delta [\epsilon - h_{\mathrm{cl}} ] + {\mathcal O}(\hbar ^{-f+1}) \nonumber \\
&+& \frac{1}{\pi \hbar}\sum_{p}\sum_{r=1}^{\infty }
\frac{\cos \left[ {r \over \hbar}S_p -
r{\pi \over 2}\nu _p\right]}{|\det ({\bf \sf m}_p^r(\lambda ) -
{\bf  \sf I})|^{1 \over 2}}\oint_pd\tau q_W(\tau )q_W(t+\tau )
\nonumber \\ &+& {\mathcal O}(\hbar ^0).
\end{eqnarray}

The propagator or frequency-dependent response function  is given by
\begin{eqnarray}
\tilde{\chi}^{\prime \prime}(\omega ) &=& {1 \over \hbar} \int dt
\int \frac{d^fxd^fp}{(2\pi \hbar )^f}
\left[p^{FD}(\epsilon )-p^{FD}(\epsilon + \frac{\hbar \omega}{2}) \right]
q_W({\bf x},{\bf p}) \nonumber \\
&~&\left[e^{(i\omega -\hat{{\mathcal L}}_{ cl})t}
q_W  \right]({\bf x},{\bf p})
+ \frac{1}{\pi \hbar ^2}
\int d\epsilon \left[p^{FD}(\epsilon )-p^{FD}(\epsilon + \frac{\hbar \omega}{2})
\right] \nonumber \\
&~&\sum_{p}\sum_{r=1}^{\infty }
\frac{\cos \left[ {r \over \hbar}S_p -
r{\pi \over 2}\nu _p\right]}{|\det ({\bf \sf m}_p^r(\lambda ) -
{\bf  \sf I})|^{1 \over 2}}
\int dte^{i\omega t}\oint_pd\tau q_W(\tau )q_W(t+\tau ).
\end{eqnarray}
Since the rate of energy dissipation is directly related to
$\tilde{\chi}^{\prime \prime}(\omega )$, the above expression gives us
a semiclassical description of dissipation. Note that the shape (which
means potential) of a system dictates the periodic orbits and the Liouvillian
operator.

\vskip 0.5 truecm

\noindent
{\bf 5. Random-matrix linear response}        \\

Energy level sequences of nuclei have local fluctuation properties
which fit remarkably well with the predictions of random matrix
theory \cite{haq}. So is the case with other chaotic quantum systems
\cite{bohigas}.
An attempt to arrive at a connection between random matrix universality
and chaos has been made recently \cite{andreev}.
In this section, we present random-
matrix expression for linear response for Fermi systems.
Thus the Fermi system is one in an    ensemble, where the Hamiltonian matrix
is modelled in terms of a random matrix.
In the context of absorption by small metallic particles, random matrix
theory was used by Gorkov and Eliashberg \cite{gorkov}.

Given a level sequence, $\{x_i\}$, the $n$-level correlation function
is defined as \cite{mehta}
\begin{eqnarray}
R_n(x_1,x_2,...,x_n) &=& \frac{N!}{(N-n)!}\int_{-\infty}^{\infty}...
\int_{-\infty}^{\infty}\nonumber \\ & &
dx_{n+1}...dx_N ~P_N(x_1,x_2,...,x_N)
\end{eqnarray}
which is the probability density of finding a level (regardless of
labelling) around each of the points $x_1,x_2,...,x_n$, the positions
of remaining levels being unobserved.
$P(x_1,...,x_N)$ denotes the joint probability distribution function
for the levels $x_1,...,x_N$.
The level density is given by the one-point function, $R_1(x)$.
The $n$-level cluster function is defined as
\begin{equation}
T_n(x_1,x_2,...,x_n) = \sum_{\mbox{ G}} (-1)^{n-m} (m-1)!
\prod_{j=1}^{m} R_{G_{j}}(x_k, \mbox{with~} x_k \in \mbox{G}_j),
\end{equation}
where G stands for any division of the indices ($1$,$2$,...,$n$) into
$m$ subgroups (G$_1$,G$_2$,...,G$_m$). For instance,
\begin{eqnarray}
&&T_1(x) = R_1(x), \nonumber \\
&&T_2(x_1,x_2) = - R_2(x_1,x_2) + R_1(x_1)R_1(x_2),
\end{eqnarray}
and so on. Since we are only interested in the local fluctuations,
the energy levels $x_i$ and $x_j$ are such that $N|x_i - x_j|~\sim
~{\mathcal O}(1)$, the so-called scaling limit. In the scaling limit,
the two-level cluster function, $T_2(x_1,x_2)$, becomes
$Y_2(|x_1-x_2|)$.

Time correlations will now be averaged over the ensemble of random
matrices which entail an ensemble-averaged time correlation function. Each term
is multiplied by the probability that two levels will be in the intervals
$d\epsilon _1$ and $d\epsilon _2$,
\begin{equation}
Y_2(|\epsilon _1 - \epsilon _2|)
\frac{d\epsilon _1}{\overline{\sigma }}
\frac{d\epsilon _2}{\overline{\sigma }}
= Y_2(|\epsilon _1 - \epsilon _2|) n_{TF}^2d\epsilon _1d\epsilon _2
\end{equation}
where $\overline{\sigma }$ is the mean level spacing and
$n_{TF} $ is the Thomas-Fermi density of levels. We get for the
response function :
\begin{eqnarray}
\tilde{\chi}^{\prime \prime} (\omega )
&=& n_{TF}^2 \sum_{m=1}^{\infty} \frac{(\hbar \omega )^m}{m!}
\int_{-\infty }^{\infty}\int_{-\infty }^{\infty}
d\epsilon _1d\epsilon _2 \frac{\partial ^m
p^{FD}(\epsilon)}{\partial \epsilon ^m}
\delta (\hbar \omega - \epsilon _2 + \epsilon _1)\nonumber \\
&&Y_2(|\epsilon _1 - \epsilon _2|)
|\langle \phi _1|\hat{q}|\phi _2\rangle |^2.
\end{eqnarray}

For the invariant ensembles, i.e., the orthogonal, the unitary, or
the symplectic ensembles, $Y_2(|\epsilon _1 - \epsilon _2|)$ embodies
universal results. Thus the random matrix expression gives the generic
form. It must be borne in mind that the local fluctuations and
universalities have been qualitatively (and sometimes quantitatively)
understood using semiclassical methods. The random matrix result provides
us a trend, a qualitative overall behaviour; it is the semiclassical theory
containing microscopic information that can eventually give us
system-specific results.

The matrix element appearing in (36) can be related to the classical
autocorrelation function of $q_{cl}$ (the classical counterpart of
$\hat{q}$) if we assume that the sum of mean values of any quantity
over all states with definite energy roughly equals the sum of the classical
mean values of this quantity as the particle moves over all the
trajectories with given energy \cite{sudhir}.

Thus,
\begin{eqnarray}
\tilde{\chi}^{\prime \prime} (\omega )
&=& \frac{n_{TF}}{2\pi } \sum_{m=1}^{\infty} \frac{(\hbar \omega )^m}{m!}
\int_{-\infty }^{\infty}\int_{-\infty }^{\infty}
d\epsilon _1d\epsilon _2 \frac{\partial ^m
p^{FD}(\epsilon)}{\partial \epsilon ^m}
\delta (\hbar \omega - \epsilon _2 + \epsilon _1)\nonumber \\
& &Y_2(|\epsilon _1 - \epsilon _2|)
\int_{-\infty}^{\infty} d\tau e^{{i \over \hbar}(\epsilon _2 - \epsilon _1)\tau}
C_{qq}(\tau )
\end{eqnarray}
where the correlation function is
\begin{equation}
C_{qq}(\tau ) = \lim _{T \rightarrow \infty} \frac{1}{T} \int_{0}^{T}
dt q(t)q(t+\tau ).
\end{equation}
Changing the variables to
\begin{eqnarray}
\epsilon _- &=& \epsilon _2 - \epsilon _1 \nonumber \\
\epsilon _+ &=& \frac{\epsilon _2 + \epsilon _1}{2}
\end{eqnarray}
and after some manipulations, we get a compact result :
\begin{equation}
\tilde{\chi}^{\prime \prime} (\omega )
= n_{TF}\frac{\hbar \omega}{2}Y_2(\hbar \omega )S_{qq}(\omega ).
\end{equation}
In this expression, $S_{qq}(\omega )$ is the spectral density defined
by
\begin{equation}
S_{qq}(\omega ) =  \frac{1}{2\pi}\int_{-\infty}^{\infty} d\tau
e^{i\omega \tau} C_{qq}(\tau ).
\end{equation}
For different Gaussian ensembles \cite{mehta} and
non-Gaussian, invariant  ensembles \cite{bipz}, the two-point function is
\begin{eqnarray}
Y_2(\hbar \omega ) &=&
\left[\frac{\sin (\pi \hbar \omega )}{
\pi \hbar \omega } \right]^2  \nonumber \\
&=&
\left[\frac{\sin (\pi \hbar \omega )}{
\pi \hbar \omega } \right]^2
+ \left[\int_{\hbar \omega }^{\infty}\left(\frac{\sin (\pi t)}{\pi t}
\right)dt \right]
\left[\frac{d}{d(\hbar \omega )}\left(\frac{\sin (\pi t)}{\pi t} \right)\right]
\nonumber \\
&=&
\left[\frac{\sin (2\pi \hbar \omega )}{
2\pi \hbar \omega } \right]^2
-\frac{d}{d(\hbar \omega )}
\left[\frac{\sin (2\pi \hbar \omega )}{
2\pi \hbar \omega } \right]\int_{0}^{\hbar \omega }
\left[\frac{\sin (2\pi t )}{ 2\pi t } \right]dt,
\end{eqnarray}
corresponding  to unitary, orthogonal, and symplectic
ensembles respectively.

For the case when a random matrix, $h$,
is perturbed by another random matrix, $q$,
independent from $h$ but invariant with respect to the same symmetry,
$S_{qq}(\omega )$ is explicitly known \cite{gjv}. The spectral density becomes
asymmetric at non-zero temperatures and shows a universal dip at zero
frequency. The behaviour of the spectral density near zero frequency
is universal, depending on the symmetry parameter.

\vskip 0.5 truecm

\noindent
{\bf 6. Response of mixed systems}        \\

The most common situation is when the dynamics is neither fully integrable
nor fully chaotic, but mixed. Looking at the phase space of such systems, we
observe stable islands embedded in stochastic sea.
We present an
expression for the response function employing the recent semiclassical result
for mixed systems \cite{creagh}.

The semiclassical expression for the density of energy levels is given by the
trace of the Green function in position representation, and it results in a sum
over periodic orbits. The integration in the method of stationary phase goes from
one point ${\bf x}$ to itself collecting contributions of periodic orbits.
Here the symmetry becomes important. If there is a continuous symmetry, then
there is a family of periodic orbits with given energy. Only the integration transverse
to this family can be done by the method of stationary phase, the parallel
integration remains. Following \cite{creagh}, let us assume that the symmetry group
is $G$. In case $G$ acts on the phase space in a way that infinitesimal
generators of the group, and the vector field of the Hamiltonian itself,
are all linearly independent, the subgroup of $G$ which leaves the periodic orbits
invariant should be discrete. In this case, the degeneracy of periodic orbits ,
$k$ is just the dimension of the group. Also, if $\gamma _0(t)$ is a reference
orbit, other orbits can be parametrised by group elements $g$ according to
\begin{equation}
\gamma _g(t) = g\gamma _0(t).
\end{equation}
The points on the family are parametrised by ($g,t$) and the $(k+1)$-dimensional
measure is $dtd\mu (g)$ where $d\mu (g)$ is the Haar measure \cite{fraleigh}.
The trace formula for the case of exact symmetry becomes \cite{creagh-littlejohn},
\begin{equation}
d(E) = \frac{1}{i\hbar} \frac{1}{(2\pi i\hbar )^{k/2}}\sum_{po, \Gamma}
\int dt d\mu (g) \mid K \mid ^{-\frac{1}{2}} e^{\frac{iS(E)}{\hbar} -
\frac{i\sigma \pi}{2}}
\end{equation}
where $K$ is an invariant  of the family and is determined by linearisation
of dynamics about a typical orbit.

For nontrivial symmetry,
\begin{equation}
K = Q\det W \det (\tilde{M} - I).
\end{equation}
Here $Q$ is the jacobian, independent of dynamics from the choice of basis for
the Lie algebra, $W$ is the symplectic matrix from the linearisation in
full phase space of the dynamics around the orbit, and tilde on $M$
represents the fact that symmetry has been extracted. In (44), $S(E)$ is
the action of a typical orbit in the family, and,
$\sigma = \mu - \delta $ where $\mu $ is the Maslov index and $\delta $ is
the number of positive eigenvalues of $W$.

Now we perturb a system (with Hamiltonian $H_0$) with continuous group
$G$ so that the resulting Hamiltonian, $H$ has no symmetry at all,
\begin{equation}
H = H_0 + \lambda H_1.
\end{equation}
For this case, the trace formula becomes \cite{creagh}
\begin{eqnarray}
d(E;\lambda ) &=& \frac{1}{i\hbar} \frac{1}{(2\pi i\hbar )^{k/2}}\sum_{po, \Gamma}
\int _{\Gamma}dt d\mu (g) \mid K \mid ^{-\frac{1}{2}} \nonumber \\
& &e^{\frac{iS_0(E)}{\hbar} -
\frac{i\sigma \pi}{2}}\langle e^{i\frac{\lambda}{\hbar}F(g,E)} \rangle
_{g \in G},
\end{eqnarray}
where
\begin{equation}
F(g,E) = - \int_{\gamma _g(t)}dt~H_1 = \frac{\Delta S}{\lambda}
({\bf x}_{\mid \mid },E;\lambda ).
\end{equation}
The angular bracket  represents an  average over the orbit label $g$.
$S_0$ is the action of a trajectory of  unperturbed Hamiltonian and
$\Delta S$ is a correction of order $\lambda $.
To provide an expression for two-time correlation function, we first
note, progressing in the same way as in Sec. 4,
\begin{eqnarray}
{\mbox tr~}\delta (E - \hat{H})\hat{X} &=&
\int \frac{d^fxd^fp}{(2\pi \hbar )^f} X_W\delta (E - H_{\mathrm{cl}}) +
{\mathcal O}(\hbar ^{-f+1})
+  \frac{1}{i\hbar} \frac{1}{(2\pi i\hbar )^{k/2}}\nonumber \\
&~&\sum_{po, \Gamma}
\int _{\Gamma}dt d\mu (g) \mid K \mid ^{-\frac{1}{2}} e^{\frac{iS_0(E)}{\hbar} -
\frac{i\sigma \pi}{2}}
\frac{\partial {\mathcal M}}{\partial \lambda}\left(\frac{\lambda}{\hbar},E\right)
\end{eqnarray}
where
\begin{equation}
{\mathcal M}\left(\frac{\lambda}{\hbar},E\right) = \langle e^{i\frac{\lambda}{\hbar}F(g,E)} \rangle
_{g \in G},
\end{equation}
and $\hat{X}$  is defined in Sec. 4.
In general, let $\hat{X}$ be $\hat{A}(0)\hat{B}(t)$.
For the case of mixed dynamics, we
obtain the following expression where the system is in thermal equilibrium
at temperature, $T$:
\begin{eqnarray}
C_{AB}(t,\lambda ,T) &=& \frac{1}{Z(\beta )}
\int \frac{d^fxd^fp}{(2\pi \hbar )^f} e^{-\beta H_{cl}}A_W(x,p)
e^{-{\mathcal L}_{cl}t}B_W(x,p) \nonumber \\ &+&
{\mathcal O}(\hbar ^{-f+1})
+  \frac{1}{i\hbar} \frac{1}{(2\pi i\hbar )^{k/2}}\nonumber \\
&~&\frac{1}{Z(\beta )}\sum_{po, \Gamma}\int dEe^{-\beta E}
\int _{t}^{t+T_{po}}dt' d\mu (g) \mid K \mid ^{-\frac{1}{2}}\nonumber \\
&~&e^{\frac{iS_0(E)}{\hbar} - \frac{i\sigma \pi}{2}}
\frac{\partial}{\partial E}\left< \exp \left[ -\frac{i\lambda}{\hbar}
\int_{\gamma _g(\tau )}d\tau ~A(\tau )B(\tau + t')\right]\right>_{g \in G}.
\end{eqnarray}

Once the group $G$ is identified, the calculation of $<>_{g \in G}$ is
not difficult. For the case of axial symmetry, it has been demonstrated
in \cite{creagh}. In the case of nuclear physics, the connection between
deformed nuclei and harmonic oscillators where the frequency ratios can
be related to deformation parameters, a much desirable simplification
is expected to occur.

Note that the decay of correlations will be governed by the eigenvalues of
the classical Liouvillian operator. For the systems which are mixing, the
decay of correlations is exponential. For mixed systems, this decay
will be more complicated and system-specific. The important point is that
dynamical aspects have been incorporated explicitly. We believe that it
will lead us to an understanding of many important aspects of many-body systems,
pre-fission neutron multiplicity, damping of giant resonances being some
of the examples.

\vskip 0.5 truecm

\noindent
{\bf 7. Irreversibility of energy dissipation} \\

It has been argued in the past that the irreversibility of energy dissipation
occurs when the single particle dynamics is chaotic \cite{hw}. Quantum mechanically,
the difficulty is in showing that the energy distribution, $\eta (E)$, defined
below follows an equation which implies irreversibility. Most often,
in literature, this
equation is taken as the Smoluchowski equation. Within the assumptions of
classical linear response, this is the correct equation, as shown by
Jarzynski \cite{jarzynski}. However, our system is quantum mechanical though we have
simplified our considerations somewhat by assuming that our system is
adiabatic. In the two subsections below, we derive evolution equations for the
energy distributions and show that these are, indeed, irreversible in time.
We assume that the single-particle dynamics is non-integrable.
It has been shown for chaotic quantum systems and also for non-integrable
systems with zero Kolmogorov-Sinai entropy that there is
level repulsion \cite{bohigas,gremaud}.
That is, the probability of clustering of two adjacent levels is zero.
In the quantum case, the equation contains a term in addition to the
Smoluchowski equation. The subsection on semiclassical adiabatic systems
shows that the quantum diffusion equation, obtained by Jain \cite{jain},
reduces to the Smoluchowski equation as $\hbar$ $\to$ $0$.

\vskip 0.5 truecm

\noindent
{\bf 7.1 Quantum adiabatic systems}\\

We consider a quantum system evolving adiabatically in time.
Since the evolution is adiabatic, there is an instantaneous basis.
Thus, the energy levels evolve in time as the system ``deforms". While
in the same symmetry class (we assume so), the levels keep evolving without
crossing. The instances when the levels come closer than the mean level
spacing, there is  non-zero probability of transitions to take place.
For instance, very small spacing between levels then leads to an
increased probability of non-adiabatic Landau-Zener transitions
\cite{zener} which  eventually, in long time,  modify the energy distribution of the
system. The basic idea behind using the Landau-Zener transitions goes
back \cite{h-w} in the literature of nuclear physics where it was used
to argue for  damping of collective modes.

Let us consider the Hamiltonian,
\begin{equation}
\hat{H}(t) = \hat{H}_0 + \epsilon t~\hat{V}
\end{equation}
where $\hat{H}_0, \hat{V}$ are linear operators.
We assume that the
evolution in time is adiabatic which corresponds to the smallness of
$\epsilon $.

At any
instant, the system admits an eigenvalue spectrum given by the eigenvalue
problem for the ``frozen" Hamiltonian,
\begin{equation}
\hat{H}(\epsilon t)|n(\epsilon t)\rangle  = E_n(\epsilon t)|n(\epsilon t)\rangle .
\end{equation}

From time $t=0$,the levels  evolve  in time, the resulting density
operator, $\hat{\rho }$ satisfies
\begin{equation}
i\hbar \frac{\partial \hat{\rho }}{\partial t} = [\hat{H},\hat{\rho }].
\end{equation}
Our objective is to derive an equation for the energy
distribution,
\begin{equation}
\eta (E) = \int^{E} dE'\mbox{tr}\{ \delta(E'-\hat{H})\hat{\rho } \}.
\end{equation}

When a system is perturbed adiabatically, there is, of course, a
clear separation of time-scales.
To incorporate these scales in the problem, we employ the multiple
time-scale method for treating the partial differential equation (54).
Accordingly, denoting the set of instantaneous states by
$\{|n(\epsilon t)\rangle \}$, we can write an expansion for the
density operator,
\begin{equation}
\hat{\rho }(\{|n(\epsilon t)\rangle \},t) = \hat{\rho }_0
(\{|n(\epsilon t)\rangle \},\epsilon t) + \epsilon \hat{\rho }_1
(\{|n(\epsilon t)\rangle \},t,\epsilon t) + ...
\end{equation}
with the initial conditions,
\begin{equation}
\hat{\rho }_0 = \hat{\rho }_{00}(\hat{H}(\epsilon t)),~\hat{\rho }_1=\hat{\rho }_2=\hat{\rho }_3=...=0.
\end{equation}
Substituting (56) in (54), we get a system of equations separated by
different orders of $\epsilon $ :
\begin{eqnarray}
[\hat{\rho }_0,\hat{H}(\epsilon t)] &=& 0, \\
i\hbar \frac{\partial \hat{\rho }_j}{\partial t} +
[\hat{\rho }_j,\hat{H}(\epsilon t)] &=& -i\hbar \frac{\partial
\hat{\rho }_{j-1}}{\partial (\epsilon t)}, ~~j=1,2,...
\end{eqnarray}
If there are no other constants of the motion than $H(\epsilon t)$ on
the fast scale, or under the Thomas-Fermi approximation, by (58),
\begin{equation}
\hat{\rho }_0(\{|n(\epsilon t)\rangle \},\epsilon t) =
\hat{\rho }_0'(H(\epsilon t),\epsilon t)
\end{equation}
where the arbitrariness of $\hat{\rho }_0'$ is removed by insisting
that $\hat{\rho }$ remains valid for times $O(\epsilon ^{-1})$ by
removing secularities in (59) with $j=1$.

To remove the secularities in
\begin{equation}
i\hbar \frac{\partial \hat{\rho }_1}{\partial t} +
[\hat{\rho }_1,\hat{H}(\epsilon t)] = -i\hbar \frac{\partial
\hat{\rho }_{0}}{\partial (\epsilon t)},
\end{equation}
in keeping with the method, we operate
by an arbitrary operator-valued function,
$g(\hat{H})$ and perform the trace of the resulting equation over
the frozen basis,
\begin{equation}
\sum_{n}\left< n\vline g\frac{\partial \hat{\rho }_1}{\partial t}
\vline n\right>  + \frac{1}{i\hbar }\sum_{n} \langle n|g[\hat{\rho }_1,
\hat{H}]|n\rangle = - \sum_{n}\left< n\vline g
\frac{\partial \hat{\rho }_0}{\partial (\epsilon t)} \vline n\right> .
\end{equation}
We should finally regard those results which hold for any $g(\hat{H})$.

For $\hat{\rho }_0$ to be valid for times for $O(\epsilon ^{-1})$,
the right hand side (RHS) of (62) should be set to zero, which leads to

\begin{equation}
\sum_{n}\left< n\vline g
\left[ \frac{\partial \rho _0'(\hat{H})}{\partial H}
\frac{\partial \hat{H}}{\partial (\epsilon t)} +
\frac{\partial \rho _0'(\hat{H})}{\partial (\epsilon t)} \right]
\vline n\right> = 0.
\end{equation}
Let us re-write the density of levels as
\begin{eqnarray}
\Sigma (E,\epsilon t) &=& \sum_{n} \langle n|\delta (E-\hat{H})|n\rangle  \nonumber \\
&=& \frac{\partial }{\partial E} \sum_{n} \langle n|\Theta (E-\hat{H})|n\rangle
= \frac{\partial \Omega (E,\epsilon t)}{\partial E}
\end{eqnarray}
where $\Omega $ is the cumulative density of levels.
In the following, it will often be  useful to convert the traces into
energy-integrals. For that,
we define an energy average over the ``frozen" Hamiltonian by
\begin{equation}
\frac{1}{\Sigma }\sum_{n} \langle n|\delta (E-\hat{H}) ... |n\rangle  =
\langle ...\rangle _{E,\epsilon t}.
\end{equation}
With this definition, it is easy to verify that
\begin{equation}
\sum_{n} \langle n|...|n\rangle  = \int dE \Sigma \langle ...\rangle _{E,\epsilon t}.
\end{equation}
Now, (63) becomes
\begin{equation}
\Sigma \left( \frac{\partial \hat{\rho }_0'}{\partial E}
\left< \frac{\partial \hat{H}}{\partial (\epsilon t)}
\right>_{E,\epsilon t} +
\frac{\partial \hat{\rho }_0'}{\partial (\epsilon t)}\right) = 0.
\end{equation}
Calling
\begin{equation}
\left< \frac{\partial \hat{H}}{\partial (\epsilon t)} \right>_{E,\epsilon t}
= u(E,\epsilon t),
\end{equation}
and using (66), we obtain the quantum adiabatic theorem:
\begin{equation}
\frac{\partial \Sigma }{\partial (\epsilon t)} + \frac{\partial }{\partial E}(\Sigma u) = 0.
\end{equation}

In the sequel, the notation $\partial _x$ stands for a partial
derivative with respect
to $x$.

This is a continuity equation for the level density. If the energy levels
are thought of as fictitious particles, the level density is like the
particle density. Indeed, in random matrix theory, one can imagine that
the levels are like particles with Coulomb interaction in two dimensions
\cite{mehta,j-a}.

Therefore, (67) reduces to
\begin{equation}
\frac{\partial }{\partial (\epsilon t)}(\hat{\rho }_0'\Sigma) +
\frac{\partial }{\partial E}(u\Sigma \hat{\rho }_0') = 0.
\end{equation}
Then, for $\hat{\rho }_0$, we have
\begin{equation}
\frac{\partial \hat{\rho }_0}{\partial (\epsilon t)}(\{|n\rangle \},\epsilon t) =
\frac{\partial \hat{\rho }_0'}{\partial E}(H,\epsilon t)
\left( \frac{\partial H}{\partial (\epsilon t)} - u \right).
\end{equation}
With this equation and the initial condition (57),
we have completely determined $\hat{\rho }_0$.

Similarly, we can proceed to determine $\hat{\rho }_1$ \cite{jain}.

Let us define the average two-time correlation function,
\begin{eqnarray}
C_{\epsilon t}(s,E)&=&
\langle \{ \partial _{{ \epsilon t}}\hat{H}(\{|n\rangle \},\epsilon t)
-u\} \{ \partial _{{ \epsilon t}}\hat{H}(\{|N\rangle \},\epsilon t)-u\}
\rangle _{{ E,\epsilon t}} \nonumber \\
&=& \frac{1}{\Sigma }\sum_{n} \biggl< n\mid \delta (E-\hat{H})\{
\partial _{{ \epsilon t}}\hat{H}(\{|n\rangle \},\epsilon t)-u\}\nonumber \\ & &
\{ \partial _{{ \epsilon t}}\hat{H}(\{|N\rangle \},\epsilon t)-u\}\mid n\biggr>,
\end{eqnarray}
and  a coarse grain over the energy spectrum, replacing thereby
\begin{eqnarray}
&& g(E_n) ~~\mbox{by} ~~ \overline{g} (E), ~\mbox{and}\nonumber \\
&& \left<n\vline \frac{\partial ^2\hat{\rho }_0'}{\partial E^2}\vline n\right>
~~ \mbox{by} ~~ \overline{\frac{\partial ^2\hat{\rho }_0'}{\partial E^2}}.
\end{eqnarray}
Also, for times of ${\mathcal O}(\frac{1}{\epsilon })$,
\begin{equation}
\int_{-t}^{0} ds C(s) = \frac{1}{2} \int_{-\infty }^{\infty } ds C(s)
= \frac{1}{2} G_2.
\end{equation}

Finally, the
condition that removes secularities to ${\mathcal O}(\epsilon ^2)$ is
\begin{equation}
\frac{\partial }{\partial (\epsilon t)}(\hat{\rho }_1'\Sigma ) +
\frac{\partial }{\partial E}(u\hat{\rho }_1'\Sigma ) -
\frac{\partial }{\partial E}\left( \Sigma G_2 \overline{\frac{\partial
\hat{\rho }_0'}{\partial E}} \right) - \frac{1}{2}\Sigma
\overline{\frac{\partial \hat{\rho }_0'}{\partial E}}\frac{\partial G_2}{\partial E}=0.
\end{equation}

The energy distribution
follows the following equation,
\begin{equation}
\frac{\partial \eta }{\partial t} = -\epsilon
\frac{\partial }{\partial E}(u\eta ) +
\epsilon ^2\frac{\partial }{\partial E}\left[
G_2\Sigma \frac{\partial }{\partial E}\left( \frac{\eta }{\Sigma }
\right)\right]
+ \frac{\epsilon ^2}{2} \Sigma \frac{\partial G_2}{\partial E}
\frac{\partial }{\partial E}\left( \frac{\eta }{\Sigma }\right),
\end{equation}
the quantum diffusion equation.
Thus, the diffusion in quantum systems has to be qualitatively
and quantitatively different as the
diffusion coefficient will be different from the one we have in the
Smoluchowski equation.

There are different important time-scales in the system.
First of all, the time scale associated with the decay of correlation function,
\begin{equation}
t_c = [C(0)]^{-1}\int_{-\infty }^{+\infty } C(s)ds.
\end{equation}
If the quantum system considered is modelled as a random matrix of dimension
$N$ \cite{gjv} with large $N$, we know that correlation function will decay
very rapidly. Thus, $t_c$ can be very small if the quantum systems possess the
following properties : (a) the number of eigenvalues is very large, and the
energy spectrum is complex, and, (b) the corresponding classical system is
chaotic. Chaos in the underlying classical system plays a fundamental role
in the decay of correlation functions. It was  shown
by Gaspard and Jain \cite{pg-sj} that
the quantum time-dependent correlations in a Fermionic system are dominated by the
classical correlation function. The decay of the correlation function is shown
in this work to be governed by the eigenvalues of the Liouvillian operator.
Thus, $t_c $ is related to the Liapunov exponents and other detailed features
of chaos. In classical ergodic adiabatic systems, the time $t$ (fast scale)
is much larger than $t_c$, thus the third term of (76) is zero. However, in
quantal systems, we have the quantum mechanical scale, $t_q=\hbar /\overline{\sigma }$
($\overline{\sigma }$ being the mean level spacing) which is why
the third term at ${\mathcal O}(\epsilon ^2)$ is explicitly present.
If $t_q \ll t_c \ll t$, the quantum effects will dominate, and all the terms
in (76) will be important. If $t_c \ll t_q \ll t$, then the system will
behave classically initially and eventually, quantum phenomena will become
important; so initial evolution will be Smoluchowski-like and then
non-Smoluchowski regime sets in. If, however, $t_c \ll t \ll t_q$, then the
evolution will be according to the classical equation. Notice, as $\hbar $
becomes small and the system is classical, $t_q$ will become
vanishingly small. In the limit, the correlations will become largely
independent of energy and the third term of (76) will drop.
We now show that the equation in the semiclassical limit is indeed the
classical Smoluchowski equation.

\vskip 0.5 truecm

\noindent
{\bf 7.2 Semiclassical adiabatic systems}\\

The combination of semiclassical and adiabatic approximations presents
a singular situation. We present here a systematic treatment
where we perform an $\hbar $-expansion in the Wigner-Weyl basis on top of
the multiple-time scale expansion. Proceeding in the same way as for the
quantum adiabatic case, we will derive an equation for the energy
distribution. The density of levels is defined as
\begin{equation}
\Sigma (E, \epsilon t) = \sum_n \langle n|\delta (E - \hat{H})|n\rangle .
\end{equation}
In the Wigner-Weyl basis, it can be written through the Wigner transform
of $\delta (E - \hat{H})$ \cite{voros},
\begin{eqnarray}
\delta (E - \hat{H})_W &=&  \delta (E - H) +
\hbar ^2 \bigg\{-\delta ^{\prime \prime}(E - H){1 \over 8}
\frac{\partial ^2V}{\partial {\bf x}^2} \nonumber \\
&+& \delta ^{\prime \prime \prime}(E - H)
{1 \over 24}\bigg[\left(\frac{\partial V}{\partial {\bf x}} \right)^2
\left({\bf p}.\frac{\partial }{\partial {\bf x}} \right)^2V
\bigg]\bigg\}.
\end{eqnarray}
Here, the Hamiltonian is assumed to be of the form ${{\bf p}^2 \over 2} + V({\bf x})$,
and the subscript $W$ refers to the Wigner transform. The Wigner transform
of $\Sigma $  is consequently written as
\begin{eqnarray}
\Sigma _W&=&\int d{\bf p} d{\bf x} \delta (E - \hat{H})_W \nonumber \\
&=&  \int d{\bf p} d{\bf x}\delta (E - H)
- \frac{\hbar ^2}{8} \int d{\bf p} d{\bf x}\delta ^{\prime \prime}(E - H)
\frac{\partial ^2V}{\partial {\bf x}^2} \nonumber \\
&-& \frac{\hbar ^2}{24} \int d{\bf p} d{\bf x}\delta ^{\prime \prime \prime}(E - H)
\bigg[\left(\frac{\partial V}{\partial {\bf x}} \right)^2
\left({\bf p}.\frac{\partial }{\partial {\bf x}} \right)^2V
\bigg].
\end{eqnarray}
The primes on the Dirac delta distribution denote the weak derivatives or
the distributionals \cite{lieb-loss}.
The energy average of an observable, $Q$, is
\begin{equation}
<Q>_{E,\epsilon t} = \frac{1}{\Sigma _W}\int d{\bf p} d{\bf x} \delta (E - \hat{H})_WQ.
\end{equation}
The Wigner transform of the density operator used in the earlier sub-section
is the Wigner distribution, $f_W$.
The evolution equation for the Wigner distribution is obtained by taking
the Wigner transform of the von Neumann equation :
\begin{eqnarray}
\frac{\partial f_W}{\partial t} &=& H \frac{2}{\hbar}\sin \bigg(
\frac{\hbar}{2}\Lambda \bigg)f_W \nonumber \\
&=& H \bigg( \Lambda - \frac{\hbar ^2}{24}\Lambda ^3 + ...)f_W.
\end{eqnarray}
Upto ${\mathcal O}(\hbar ^2)$, for the Hamiltonian of the form specified above,
the evolution equation for $f_W$ is
\begin{equation}
\frac{\partial f_W}{\partial t} =  \frac{\partial V}{\partial {\bf x}}.
\frac{\partial f_W}{\partial {\bf p} } - {\bf p}.\frac{\partial f_W}{\partial {\bf x}}
-\frac{\hbar ^2}{24} \frac{\partial ^3V}{\partial {\bf x}^3}.
\frac{\partial ^3f_W}{\partial {\bf p}^3}.
\end{equation}
We now expand $f_W$ in a power series of $\hbar$ :
\begin{equation}
f_W = f^{(0)} + \hbar ^2f^{(2)} + \hbar ^4f^{(4)} + ...
\end{equation}
As a result, we obtain the following hierarchy of equations :
\begin{eqnarray}
{\mathcal O}(\hbar ^0)&:& ~~~~~\partial _t f^{(0)} = \{H,f^{(0)}\}, \nonumber \\
{\mathcal O}(\hbar ^2)&:& ~~~~~\partial _t f^{(2)}
= \{H,f^{(2)}\} -\frac{1}{24}\partial _{\bf x}^3V.\partial _{\bf p}^3f^{(0)}, \nonumber \\
\vdots \nonumber \\
{\mathcal O}(\hbar ^{2n})&:& ~~~~~\partial _t f^{(2n)}
= \{H,f^{(2n)}\} -\frac{1}{24}\partial _{\bf x}^3V.\partial _{\bf p}^3f^{(2(n-1))}.
\end{eqnarray}
$\{.,.\}$ denotes the Poisson bracket in the mock phase space variables.
To solve these equations, we shall exploit the two time-scales, $t$ and
$\epsilon t$, and expand $f^{(0)}$ ($({\bf x},{\bf p})={\bf z}~ ~\in ~~$mock phase space
\cite{balasz}) as
\begin{equation}
f^{(0)} = f_0^{(0)}({\bf z},\epsilon t) + \epsilon f_1^{(0)}({\bf z},\epsilon t,t) +
\epsilon ^2f_2^{(0)}({\bf z},\epsilon t,t) + ...
\end{equation}
with the initial condition,
\begin{equation}
f_0^{(0)}({\bf z},0) = f_{00}, ~~f_i^{(0)} = 0 ~~\forall ~~i\geq 1 ~~\mbox{at~~} t = 0.
\end{equation}
To various orders in $\epsilon $, we have a hierarchy of equations
nested in the $\hbar $-hierarchy:
\begin{eqnarray}
{\mathcal O}(\epsilon ^0) &:& ~~~~~\{f_0^{(0)}, H\} = 0, \nonumber \\
{\mathcal O}(\epsilon ^n)&:& ~~~~~\frac{\partial f_n^{(0)}}{\partial t}
+ \{f_n^{(0)}, H\} = -\frac{\partial f_{n-1}^{(0)}}{\partial (\epsilon t)}.
\end{eqnarray}
Due to the ${\mathcal O}(\epsilon ^0)$ equation,
\begin{equation}
f_0^{(0)}({\bf z},\epsilon t) = f_0(H,\epsilon t )
\end{equation}
with the initial condition,
\begin{equation}
f_0(E,0) = f_{00}(E).
\end{equation}
$f_0$ is completely specified by (90) and
\begin{equation}
\frac{\partial}{\partial (\epsilon t)}(\Sigma f_0) +
\frac{\partial }{\partial E}(u\Sigma f_0) = 0
\end{equation}
removing secularities at ${\mathcal O}({1 \over \epsilon})$ in time.
In (91), we have employed the following notations :
\begin{eqnarray}
\Sigma (E,\epsilon t) &=& \int d{\bf z} \delta (E-H) =
\frac{\partial}{\partial E}\Omega (E, \epsilon t), \nonumber \\
<...>_{E,\epsilon t} &=& \frac{1}{\Sigma }\int d{\bf z} \delta (E-H)..., \nonumber \\
u &=& \left< \frac{\partial H}{\partial (\epsilon t)}\right> _{E,\epsilon t}.
\end{eqnarray}
Eqs. (88), (91), and (92) imply
\begin{equation}
\frac{\partial f_0^{(0)}}{\partial (\epsilon t)}({\bf z},\epsilon t)
= \frac{\partial f_0}{\partial E}(H,\epsilon t)
\left(\frac{\partial H}{\partial (\epsilon t)}
- u(H, \epsilon t)\right).
\end{equation}
The formal solution of (88) for $n=1$ is
\begin{equation}
f_1^{(0)}({\bf z},t,\epsilon t) = - \int_0^t dt'
\frac{\partial f_0^{(0)}}{\partial (\epsilon t)}({\bf Z},\epsilon t)
+ f_1(H, \epsilon t),
\end{equation}
where ${\bf Z} = {\bf Z}({\bf z},t,t',\epsilon t)$ is a point in mock phase space reached
by starting at ${\bf z}$ at time $t$, then evolving backward in time to $t'$
under $H$.

We can determine $f_1^{(0)}$ by removing secularities at
${\mathcal O}(\epsilon ^{-2}$) of time. Before doing so, we need some
definitions which will be encountered in the sequel. The two-time
correlation function is
\begin{equation}
C(s) = \left< \left( \frac{\partial H}{\partial (\epsilon t)} - u\right)
{\sf O}_{\epsilon t}(s)
\left( \frac{\partial H}{\partial (\epsilon t)} - u\right)\right>,
\end{equation}
where the operator ${\sf O}$ acts to its right,
evolving a point ${\bf z}$ for a time $s$ under $H$.
$C(s)$ satisfies the same conditions as in the previous sub-section.
The condition avoiding secularities at ${\mathcal O}(\epsilon ^2)$ is
\begin{equation}
\frac{\partial}{\partial (\epsilon t)}(\Sigma f_1) +
\frac{\partial}{\partial E}(u\Sigma f_1) -
{1 \over 2}\frac{\partial}{\partial E}
\left(\Sigma G_2 \frac{\partial f_0}{\partial E}\right) = 0.
\end{equation}
Thus, valid to ${\mathcal O}(\epsilon )$, the distribution function is
\begin{eqnarray}
f^{(0)}({\bf z}, \epsilon t) &=& f_0(H, \epsilon t) + \epsilon
f_1(H, \epsilon t) \nonumber \\
&-&\epsilon \frac{\partial f_0}{\partial E}(H, \epsilon t) \int_{0}^{t} dt'
\left(\frac{\partial H}{\partial (\epsilon t)}({\bf Z}, \epsilon t)
- u \right),
\end{eqnarray}
where $f_0$ and $f_1$ satisfy (91) and (96) respectively.

Now we expand the quantum correction, $f^{(2)}$, as $f^{(0)}$ with
$\hbar $ and $\epsilon $ as independent parameters :
\begin{equation}
f^{(2)} = f_0^{(2)}({\bf z}, \epsilon t) + \epsilon
f_1^{(2)}({\bf z}, \epsilon t, t) + \epsilon ^2 f_2^{(2)}({\bf z}, \epsilon t, t) + ...
\end{equation}
To orders in $\epsilon $, we have
\begin{eqnarray}
{\mathcal O}(1) &:& \{H, f_0^{(2)}\} = \frac{1}{24} \partial _{\bf x}^3.
\partial _{\bf p}^3 f_0^{(0)}, \nonumber \\
{\mathcal O}(\epsilon ^i) &:& \frac{\partial f_i^{(2)}}{\partial t} =
\{H, f_i^{(2)}\} - \frac{\partial f_{i-1}^{(2)}}{\partial (\epsilon t)}
- \frac{1}{24} \partial _{\bf x}^3.\partial _{\bf p}^3f_i^{(0)}, ...
\end{eqnarray}
The solution of (99) (for $f_0^{(2)}$) is \cite{smerzi}
\begin{equation}
f_0^{(2)}({\bf z}, \epsilon t) = \frac{1}{8} \left\{
-\partial _{\bf x}^2V \partial _E^2f_0^{(0)} - \left[
\frac{1}{3} (\partial _{\bf x}V)^2 + \frac{1}{3} ({\bf p}.\partial _{\bf x})^2V({\bf x})\right]
\partial _E^3f_0^{(0)}\right\}.
\end{equation}
Upto ${\mathcal O}(\epsilon )$ and ${\mathcal O}(\hbar ^2)$, the  semiclassical
density is
\begin{eqnarray}
f_W({\bf z},\epsilon t, t) &=& f_0(h, \epsilon t) + \epsilon
f_1(h, \epsilon t) - \epsilon \frac{\partial f_0}{\partial E}(h, \epsilon t)
\nonumber \\ & &
\int_{0}^{t} dt' \left[\frac{\partial h}{\partial (\epsilon t)}(Z, \epsilon t)
- u(h, \epsilon t) \right] +
\frac{\hbar ^2}{8} \biggl\{
-\partial _{\bf x}^2V \partial _E^2f_0^{(0)} \nonumber \\&-&
\biggl[ \frac{1}{3} (\partial _{\bf x}V)^2
+ \frac{1}{3} ({\bf p}.{\bf \partial _{\bf x}})^2
V({\bf x})\biggr]
\partial _E^3f_0^{(0)}\biggr\}.
\end{eqnarray}
Now we consider (99). Multiplying by a test function, $g(h)$ in the
equation for $i=1$ and integrating over the phase space gives us
\begin{equation}
\frac{\partial }{\partial t} \int d{\bf z} g(h)f_1^{(2)}
= -\int {\bf z} g(h)\frac{\partial f_0^{(2)}}{\partial (\epsilon t)}
- \frac{1}{24} \int d{\bf z} g(h)\partial _{\bf x}^3V.\partial _{\bf p}^3f_1^{(0)}.
\end{equation}
Since the RHS is independent of $t$, the LHS grows secularly. We set the
RHS to zero, and after some manipulations, we get
\begin{equation}
\frac{\partial }{\partial (\epsilon t)}(\Sigma f_0^{(2)})
+ \frac{\partial }{\partial E} (u\Sigma f_0^{(2)})
+ \int dE \Sigma \left< \frac{1}{24}\partial _{\bf x}^3V.
\partial _{\bf p}^3f_1^{(0)}\right>_{E, \epsilon t} = 0.
\end{equation}
With (91), (96), and (103), we obtain the semiclassical diffusion
equation,
\begin{eqnarray}
\frac{\partial \eta }{\partial t} = &-&\epsilon \frac{\partial }{\partial E}
(u\eta ) + {1 \over 2}\epsilon ^2\frac{\partial }{\partial E}
\left[ G_2 \Sigma \frac{\partial }{\partial E}\left(\frac{\eta }{\Sigma }
\right) \right]\nonumber \\
&-& \hbar ^2 \int dE \Sigma (E, \epsilon t)
\left< \frac{1}{24}\partial _{\bf x}^3V.
\partial _{\bf p}^3f_1^{(0)}\right>_{E, \epsilon t}.
\end{eqnarray}
We have not succeeded to include $f_1^{(2)}$ in this work.

Note that (104) reduces to the Smoluchowski equation as
$\hbar \rightarrow 0$.

\vskip 0.5 truecm

\noindent
{\bf 8. Geometric phase and dissipation}\\

Adiabatic approximation leads to linear response theory on one hand
where dynamical susceptibility (or polarization propagator) is central,
and geometric phases on the other. When a particle (e.g., a nucleon)
moves inside an enclosure whose boundary is adiabatically vibrating in
time, the wavefunction can acquire a geometric phase over a cycle of
vibration. In this section, following \cite{jain-pati}, we show that
the geometric phase thus acquired is related to the imaginary part of
susceptibility. It is well-known that the imaginary part of susceptibility
is a quantifier of dissipation, geometric phase plays an important role
in understanding of damping of collective excitations.

To understand how this relation comes about, for the sake of brevity, we
restrict ourselves to the case of cyclic evolution of deformations of the
many-body system. Let us denote the Hamiltonian parametrised by some
parameters (e.g., deformation parameters), ${\bf R}$, by $H({\bf R})$ which
describes a single particle in an effective mean-field. It is well-known
\cite{berryph} that when the parameters evolve along a cyclic path ${\mathcal C}$
the instantaneous eigenfunction of the system $\mid n({\bf R})\rangle $
corresponding to the eigenvalue, $E_n({\bf R})$, acquires a geometric phase
given by
\begin{eqnarray}
\gamma _n({\mathcal C}) &=& \oint_{{\mathcal C}} i\langle n({\bf R})\mid \nabla _
{{\bf R}}n({\bf R})\rangle .d{\bf R} \nonumber \\
&=& -\frac{1}{\hbar}\int_{{\mathcal S}}{\bf V}_n.d{\bf S},
\end{eqnarray}
where ${\mathcal S}$ is the surface enclosed by ${\mathcal C}$ in the parameter
space, and ${\bf V}_n$ is the ``field strength" (adiabatic curvature) given
by a familiar expression involving a wedge product :
\begin{equation}
{\bf V}_n = -i\hbar \sum_{m(\neq n)}\frac{\langle  n|\nabla _{\bf R}H|m\rangle
\wedge \langle  n|\nabla _{\bf R}H|m\rangle }{(E_n-E_m)^2}.
\end{equation}
A suitable form of ${\bf V}_n$ for the sequel is \cite{jr}
\begin{equation}
{\bf V}_n = \frac{i}{2\hbar } \lim_{\epsilon \rightarrow 0} \int_{0}^{\infty}
 dt e^{-\epsilon t} t \langle  n|\left[(\nabla _{\bf R}H)_t,\wedge (\nabla _{\bf R}H)
 \right]|n\rangle
\end{equation}
where $(\nabla _{\bf R}H)_t$ denotes the Heisenberg-evolved operator.
Note that, the state $|n({\bf R})\rangle $ appearing in  (105) corresponds to a single-particle
eigenket in an effective mean-field. This state is clearly related to the
original many-body Fermi system for which the imaginary part of the
dynamical susceptibility is \cite{kubo,balescu}
\begin{equation}
\chi ^{\prime \prime}(t) = \frac{1}{2\hbar }\langle  \Phi_0|[\hat{\mathcal A}(t),\hat{\mathcal B}(0)]|\Phi_0 \rangle
= \int \frac{d\omega }{2\pi }e^{-i\omega t}\tilde{\chi }^{\prime \prime}(\omega )
\end{equation}
where $|\Phi _0\rangle $ is the pure ground state of the many-particle system
with Fermi energy, $E_F$.
If one-body operators, $\hat{H}$, $\hat{A}$, and $\hat{B}$,
are  used  to construct many-body operators by a direct sum so as
to get
$\hat{\mathcal    H}$,   $\hat{\mathcal   A}$,   and    $\hat{\mathcal   B}$,
respectively,
and
$\hat{\mathcal H}|\Phi _l\rangle ={\mathcal E}_l|\Phi _l\rangle $, then we \cite{pg-sj}
have
\begin{equation}
\langle \Phi _0|\hat{\mathcal A}|\Phi _l\rangle  = \langle m({\bf R})|\hat{A}|n({\bf R})\rangle .
\end{equation}

On reducing the many-body system at $T = 0 K$, (where the Fermi-Dirac distribution
is a Heaviside step function), to one-body system, we can express \cite{pg-sj}
\begin{equation}
\tilde{\chi }^{\prime \prime}_{AB}(\omega  ) = -\frac{\omega }{2}\int  dt
e^{i\omega t}~\mbox{tr~}\delta (E_F-H)[\hat{A}(t),\hat{B}(0)].
\end{equation}
This can be written as in Sec. 4 semiclassically.

The label $n$ in (105) corresponds to single-particle states and is related to
$|\Phi _0\rangle $ because the many-body matrix elements can be written in terms of
one-body matrix elements for the case when all the constituents are
taken as non-interacting. In many-body physics, this gives the zero-order
response whereupon the interaction can be included in a Vlasov description
in an iterative way \cite{brink}. For relating the response function to the
geometric phase, the operators $\hat{\mathcal A}$ and
$\hat{\mathcal B}$ in our discussion are to be identified with $\nabla _X\hat{\mathcal H}$ and
$\nabla _Y\hat{\mathcal H}$ for ${\bf R}=(X, Y, Z)$.

The matrix element in (105) can be written as a many-body matrix element
using (109) by composing $\hat{A}$ and $\hat{B}$ so that we get the operator,
${\mathcal C}(t)=[{\mathcal A}(t),{\mathcal B}(0)]-[{\mathcal B}(t),{\mathcal A}(0)]$,
which is related to a difference $\chi ^{\prime \prime }_{AB}(t)-
\chi ^{\prime \prime } _{BA}(t)=\chi ^{\prime \prime }_C(t)$.
Thus, we can re-write ${\bf V}_n$ as
\begin{eqnarray}
{\bf V}_n &=& \frac{i}{2\hbar } \lim_{\epsilon \rightarrow 0} \int_{0}^{\infty }
dt e^{-\epsilon t}~t~\langle \Phi _0|{\mathcal C}|\Phi _0 \rangle \nonumber \\
&=& \int_{0}^{\infty } dt~t\chi ^{\prime \prime }_C(t)
= -\frac{\partial \tilde{\chi }^{\prime \prime }(\omega)}{\partial \omega }\vline_{\omega = 0}.
\end{eqnarray}
We now arrive at our  relation for the case of cyclic evolution :
\begin {equation}
\gamma _n({\mathcal C}) = \int_{\mathcal S} d{\bf S}.
\frac{\partial \tilde{\chi }^{\prime \prime}_{C}(\omega ;{\bf R})}{\partial \omega }
\vline_{\omega =0}.
\end{equation}
Since $\tilde{\chi }^{\prime \prime }_{C}(\omega )$ is related to energy dissipation,
this relation connects geometric phase to dissipation. In other words, we see
that dissipation in finite systems is  purely quantum mechanical as it is
related to geometric phase. That this forms a fundamental basis for the
Bohr's liquid drop model is discussed elsewhere \cite{jain-pati}.

Similarly, we can find a relation for the case of non-cyclic evolutions.

It is known that friction or viscosity in a quantal system can appear
in a thermodynamic limit. Since we have a many-body system, seemingly
contradictory conditions of finiteness of the system and ``continuity"
of the energy spectrum are met with. The finite size explicitly manifests
itself in terms of a sum over periodic orbits.

It is important to note that the wavefunction that we have considered
above is the one which satisfies the requirements in Section 3.

It is interesting to note that the viscosity of quantum Hall fluid in
two dimensions at zero temperature is related to adiabatic curvature
\cite{avron}. Also, for covalent dielectrics, the polarisation is
related to geometric phase. These two examples and our general treatment
leads us to think of a deeper unified connection.

\vskip 0.5 truecm

\noindent
{\bf 9. Fission viscosity tensor}\\

Since long, linear response theory is used to study the dynamics of
fission process \cite{jensen} in nuclei. The collective coordinates are the
deformation parameters  and pairing gap. The quantity related to fission
viscosity tensor is the first moment of the time-dependent response function.
In this paper, we have presented  semiclassical and random matrix  expressions
for the response function. In this Section, we apply this knowledge to
obtain expression for viscosity tensor in terms of periodic orbits of the
underlying classical system. It will also turn out  that
the classical Liouvillian operator plays a fundamental role.
This is very interesting as the eigenvalues of the Liouvillian
dictate the decay of correlations. Thus, the existence of dissipation
is directly related to the non-integrability of the classical system.

If we denote by $\{q_{\nu }\}$ the set of $N$ collective coordinates, then
the friction viscosity tensor is given by
\begin{equation}
\gamma _{\nu \mu } = ~^{(1)}M_{\nu \mu },
\end{equation}
where $ ~^{(n)}M_{\nu \mu }$ are the moments of the
time-dependent response function, $C_{\nu \mu }(t,T)$. The system is at
a temperature, $T$, related to the excitation energy of the intrinsic
system. It was brought into the calculations when the intrinsic degrees
of freedom were averaged over the canonical ensemble. Once again, we recall
that the canonical statistical mechanics follows from the assumption of
microscopic chaos in single-particle motion.

The semiclassical form of the time-dependent response function
(or the time correlation function) gives the following expresssion for
the viscosity tensor when we assume that the single-particle-dynamics
is fully chaotic (in case the dynamics is mixed, expression can be
immediately written using (51)) :
\begin{eqnarray}
\gamma _{\nu \mu } &=& \int_{0}^{\infty } dt~t C_{\nu \mu }(t,T) \nonumber \\
&=& \frac{1}{Z(\beta )}\int \frac{d^fxd^fp}{(2\pi \hbar )^f}
e^{-\beta H_{\mathrm cl}} \int_{0}^{\infty } dt~t
\left( \frac{\partial H}{\partial q_{\nu }}\right)_W
\exp (-\hat{{\mathcal L}}_{\mathrm  cl})
\left( \frac{\partial H}{\partial q_{\mu }}\right)_W \nonumber \\
&+& \frac{1}{\pi \hbar Z(\beta )}\sum_{p}\sum_{r=1}^{\infty }
\int dE e^{-\beta E} \frac{\cos\left(\frac{r}{\hbar}S_p(E)-r\frac{
\pi}{2}\nu _p\right)}{\mid \det ({\sf m}_p^r(E) - {\sf I})\mid ^{1/2}}\nonumber \\
&~& \int_{0}^{\infty }dt~t\oint_p d\tau
\left(\frac{\partial H}{\partial q_{\nu }}\right)_W(\tau )
\left(\frac{\partial H}{\partial q_{\mu }}\right)_W(\tau + t).
\end{eqnarray}

This is a semiclassical expression based on incorporation of
the Gutzwiller trace formula in the linear response theory. Applications
to the practical situations in nuclear fission will be of a great interest.

A random matrix expression for the viscosity tensor can also be readily
written employing the results of Section 5. The coefficient of friction
in the zero-frequency limit is
\begin{equation}
\gamma _0 = -\frac{\partial \tilde{\chi}^{\prime \prime}(\omega )}{\partial
\omega}~\vline _{\omega = 0}
\end{equation}
where $\tilde{\chi}^{\prime \prime}(\omega )$ is given by (40). Notice that
friction is physically the adiabatic curvature, from (112).
This has been also found earlier
for  viscosity of  quantum Hall systems \cite{avron}.

\vskip 0.5 truecm

\noindent
{\bf 10. Summary}\\

A lot of work has been
done on the nature of nuclear dissipation (see, e.g., \cite{ivanyuk}).
It has been stated, most clearly by Weidenm\"{u}ller \cite{hw}, that chaos in the
nucleonic motion will be related to dissipation.
However, before trying to explain the experimental data, we need a
systematic theory.
In this work, we have
taken a step in realising this connection.

On the basis of the connection established between quantum chaos and
statistical mechanics, we have developed semiclassical and random matrix
theory of response to  external perturbations of a many-body Fermi system.
For the cases where  the dynamics of a single particle is chaotic or not
fully chaotic but mixed, we
have presented the results for response functions.
Also, we have shown that the collective
excitations damp irreversibly and this irreversibility of damping is shown
to be quantum mechanical.
The damping of collective motion is reflected in the re-distribution of
energy. Energy re-distribution has been shown to evolve irreversibly.
The reason for this irreversibility is
the complexity of the energy spectrum which might, in turn, be connected with
classical non-integrability (chaos or mixed dynamics). The complexity is
such that, in the scaling limit of the energy spectrum,
the system behaves as if it is a realisation of
an ensemble of random matrices. Quantum chaos or complexity of spectrum
brings with itself a kind of almost periodic evolution so that the system
takes much longer to show any recurrence. The recurrence is a must for a
system with a discrete spectrum. However, collective excitations decay much
before this recurrence can occur. To present a simple
instructive example,  we consider
the set of prime numbers and look at
\begin{equation}
{\mathcal P}(t) = \sum_{\mathrm{primes }, p}~~\exp [ipt],
\end{equation}
which can be thought of as the survival probability (a special case of
two-time correlation function) of a fictitious system.
This sum decays quickly as the number of primes increase. Clearly then,
for a many-body sytem like a nucleus where  the spectrum is quite complex,  the
time correlations decay very quickly.
This will be decided by the spectral properties of the Liouvillian \cite{gaspard}.
The coarse-graining we had done in Sec. 7 was  a
representative of the complexity we have discussed here.

Another point of
importance is the arbitrariness of the initial density operator. We have
obtained quantum and semiclassical generalisations of the Smoluchowski
equation. Damping is described
in terms of the imaginary part of the response function. This is    shown
to be related to geometric phase acquired by a single-particle wavefunction
as the many-body system evolves.

The application of the general theory based on existence of chaos or mixed
chaos in the many-body system is being applied to fission of nuclei and
metallic clusters. We have presented an explicit expression for the fission
viscosity tensor in Sec. 9 which can be used to analyse the fission data for relatively
heavier nuclei.
Clearly, the quantum diffusion equations obtained here
are directly applicable to fission processes \cite{thoe,bjornholm} where
the Smoluchowski equation has been used until now \cite{hans}.
The difference new term will make in the assignment of transport coefficients
is of interest.
In case of nuclei, in the past \cite{pal-ganguly}, there has been a
criticism in modelling the deforming mean-field by the form,
$\hat{Q}F^{\mathrm{ext}}(t)$. However, this is
basically a matter of convenience.
However, while making realistic calculations, one may keep this in mind.
The advantage one has here is that
one gets the response function.

With the semiclassical wall formula obtained  here,
we are now trying to implement
it by finding periodic orbits of the underlying classical system. Since the
trace formulae can be written in terms of zeta functions, the numerical
implementation will be efficient. Hopefully, not too many periodic orbits will
be needed. In this regard, since the deformed nuclei can be modelled in terms
of harmonic oscillator potentials, and we have the exact trace formulae for
them \cite{brack-jain}, it will be interesting to use our expressions to find
response functions.

We hope to study nonlinear response in a semiclassical and random-matrix
setting in future. This is important for understanding phenomena at an
ultra-fast time scale.

Hence, we have shown that there is
an emergence of a new understanding of damping of collective excitations,
on quantum transport coefficients in finite Fermi systems, and in the general
theory of response of nuclei, metallic clusters, and quantum dots.

\newpage

\noindent
{\Large \sf Acknowledgements}

\vskip 0.3 in

The author expresses profound gratitude to
Dipak Chakravarty and Vivek Datar for numerous discussions on many
aspects of this work.


\begin{thebibliography}{99}
\bibitem{footnote2} It is interesting to note that the equation we obtain
for quantum diffusion turns out to be the same as we generally encounter
in weak-turbulence plasma theory.
\end{thebibliography}
\end{document}